# Interaction of topological solitons with defects: using a nontrivial metric

**Kurosh Javidan**

Department of physics, Ferdowsi university of Mashhad, 91775-1436 Mashhad Iran

javidan@um.ac.ir

**Abstract**
By including a potential into the flat metric, we study the interaction of sine-Gordon soliton with different potentials. We will show numerically that while the soliton-barrier system shows fully classical behaviour, the soliton-well system demonstrates non-classical behaviour. In particular, solitons with low velocities are trapped in the well and radiate energy. Also for narrow windows of initial velocity, soliton reflects back from a potential well.

## 1. INTRODUCTION

Toplogical solitons are important objects in nonlinear field theories. They are stable against dispersive effects, and "live" similar to classical point-like particles.
Scattering of solitons from potentials (which generally come from medium properties) have been studied in many papers by different methods [1-7]. The effects of medium disorders and impurities can be added to the equation of motion as perturbative terms [1, 2]. These effects also can be taken into account by making some parameters of the equation of motion to be function of space or time [3, 4]. There still exists another interesting method which is mainly suitable for working with topological solitons [5, 6]. In this method, one can add such effects to the Lagrangian of the system by introducing a suitable nontrivial metric for the back ground space-time, without missing the topological boundary conditions [5-7]. In other words, the metric carries the information of the medium. By adding the potential to the metric, the total energy of "soliton + potential" and also topological charge of the soliton will remain conserved during the soliton-potential interaction [7]. This method can be used for objects that their equation of motion results from a Lorentz invariant action, such as sine-Gordon model, $\varphi^4$ theory, $CP^N$ model, Skyrme model, Faddeev-Hopf equation, chiral quark-soliton model, Gross-Neveu model, nonlinear Klein-Gordon models and so on. In this paper, we have used a nontrivial metric for coupling the kink of sin-Gordon model to a potential. We find that the method is very powerful and several important properties of soliton-potential interactions come out easily.

It was pointed out in [9] that the scattering of non-topological solitons from defects show some non-classical properties at the low velocities. Motivated by this, it is natural to search for non-classical behaviour in topological solitons. Baby Skyrme model which contains topological solitons has been used for searching such behaviour [10]. In [10], the potential barrier and well has been simulated by adding an extra field to the Lagrangian. In the present paper, however we are interested in the method of adding the potential through the metric [6, 7]. Using this method, we study numerically the scattering of sine-Gordon solitons with the potential and search for the non-classical behaviours.

## 2. Potentials and the metric

Space dependent potentials can be added to the Lagrangian of a system, through the metric of background space-time. So the metric includes characteristics of the medium. The general form of the action in an arbitrary metric is:

$$I = \int \ell(\varphi, \partial_\mu \varphi)\sqrt{-g}\, d^n x\, dt \quad (1)$$

Where "g" is determinant of the metric $g^{\mu\nu}(x)$. Energy density of the "field + potential" can be found by varying "both" the field and the metric (See [7]). Simulations show that the "total" energy is conserved during the field-potential interaction. Here we are interested in the evolution of energy of the field itself. So we have to calculate the soliton energy, by varying only the field, in (1).

Sine-Gordon model is a well-known equation which contains topological solitons. Lagrangian of the sine-Gordon model is:

$$\ell = \frac{1}{2}\partial^\mu \varphi \partial_\mu \varphi - U(\varphi) \quad (2)$$

With $U(\varphi) = 1 - \cos\varphi$.

The equation of motion for the Lagrangian (2) is [8]:

$$\frac{1}{\sqrt{-g}}\left(\sqrt{-g}\,\partial_\mu \partial^\mu \varphi + \partial_\mu \varphi \partial^\mu(\sqrt{-g})\right) + \frac{\partial U(\varphi)}{\partial \varphi} = 0 \quad (3)$$

The suitable metric in the presence of a weak potential V(x) is [5, 6, and 7]:

$$g_{\mu\nu}(x) \cong \begin{pmatrix} 1+V(x) & 0 \\ 0 & -1 \end{pmatrix} \quad (4)$$

The equation of motion (3) (describes by Lagrangian (2)) in the background space-time (4), is [8]:

$$(1+V(x))\frac{\partial^2 \varphi}{\partial t^2} - \frac{\partial^2 \varphi}{\partial x^2} - \frac{1}{2|1+V(x)|}\frac{\partial V(x)}{\partial x}\frac{\partial \varphi}{\partial x} \\ + \frac{\partial U(\varphi)}{\partial \varphi} = 0 \quad (5)$$

The field energy density is:

$$\aleph = g^{00}(x)\left(\frac{1}{2}g^{00}(x)\dot{\varphi}^2 + \frac{1}{2}\varphi'^2 + U(\varphi)\right) \quad (6)$$

and the topological charge density is:

$$Q = \int_{-\infty}^{\infty} \frac{\partial \varphi}{\partial x}dx. \quad (7)$$

Equation (7) shows that topological charge is independent of the metric and consequently from the potential.

Localized solutions of the sine-Gordon equation in flat space time are:

$$\varphi(x) = 4\tan^{-1}\left(\pm \exp\left(\frac{(x-x_0)-ut}{\sqrt{1-u^2}}\right)\right) \quad (8)$$

The plus (minus) sign is soliton (anti soliton) which moves with speed "u" from the initial place $x_0$. We can use these solutions as initial condition for solving (5), if soliton is located far from the center of the potential. A potential of the form of $V(x) = ae^{-b(x-c)^2}$ has been chosen in the simulations. Parameter "a" controls the strength of the potential, "b" represents its range, and "c" indicates center of the potential. If $a > 0$, the potential shows a barrier and for $a < 0$ the potential shows a well. I have performed simulations using 4th order Runge-Kutta method for time derivatives. Space derivatives were expanded by using finite difference method. I have used grid spacing h=0.01, 0.02 and sometimes h=0.001. Time step has been chosen as $\frac{1}{4}$ of the space step "h". Simulations have been setup with fixed boundary conditions and solitons have been kept far from the boundaries. We have controlled the results of simulations by checking the conserved quantities: total energy and topological charge, during the evolution. It is clear that the energy of soliton "itself" changes in time. But the energy of soliton + potential remains unchanged. Here we interested in the energy of soliton and its evolution during the interaction. If we subtract the soliton energy density from total energy density, we find the shape of the potential. Also we can place a static soliton at different places and calculate its total energy. This tells us what the potential is like as seen by the soliton [10].

### 3. Interaction of a Soliton with a barrier

Consider the potential $V(x) = a\exp(-b(x-c)^2)$ with $a > 0$. Figure 1 shows the effective potential as seen by the soliton (solid line). The dashed line in figure 1 shows $E_0 + V(x)$, where $E_0$ is the energy of a static kink far from the potential and V(x) is the potential function. This figure is the result of using the interesting method which was explained in the end of the previous section. Figure 1 demonstrates the similarities and the differences between V(x) and the effective potential.

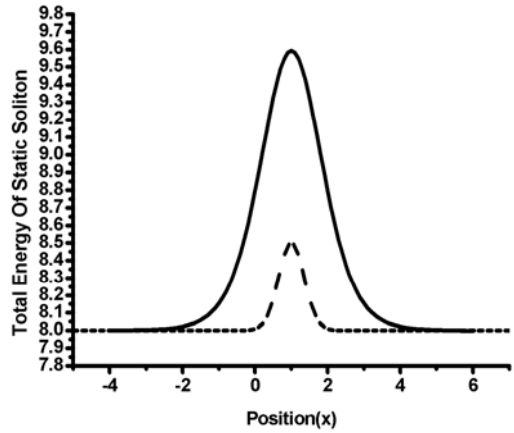

**Figure 1:** Total energy of a static soliton (solid line) placed in different position, under the influence of the potential $V(x) = 0.5\exp(-4(x-1)^2)$ and $E_0 + V(x)$ (dashed line).

Suppose a soliton is placed far away from the potential, with different initial velocities. It moves toward the barrier and interacts with it. Numerical simulations show us details of the interaction.
In a set of simulations I chose a soliton placed in x=-14 with different initial velocities and studied its interaction with the potential $V(x) = 0.5\exp(-4(x-1)^2)$ looking for the non-classical behaviours, like bound states. There exist two different kinds of trajectories for the soliton during the interaction by the barrier, depend on its initial velocity, which separate by a critical velocity $u_c$. In low velocities $u_i < u_c$, soliton reflects back and reaches its initial place with final velocity $u_f = -u_i$. Figure 2 presents trajectories of a soliton with different values of $u_i < u_c$.

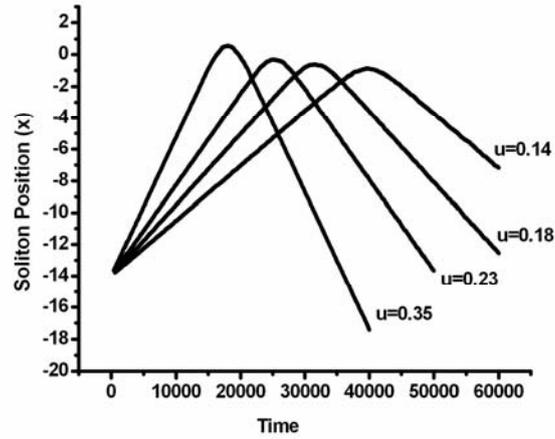

**Figure 2**: Trajectory of soliton with different values for $u_i < u_c$.

The soliton velocity is calculated by numerical differentiation of its trajectory with respect to time steps. Figure 3 shows the soliton velocity with initial speed of u=0.16, during the interaction with the potential $V(x) = 0.5\exp(-4(x-1)^2)$.

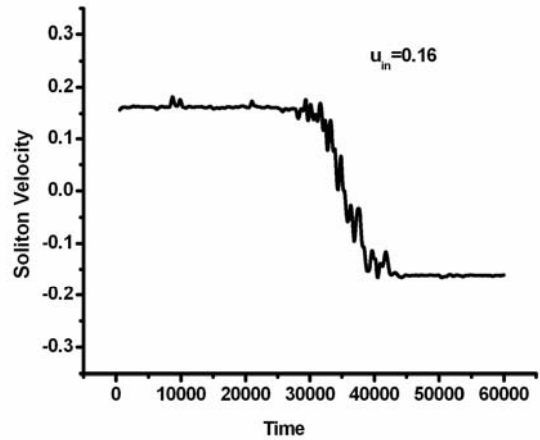

**Figure 3:** Soliton velocity during the interaction with potential $V(x) = 0.5\exp(-4(x-1)^2)$.

As we can see in figure 3, the final speed is equal to the initial speed which shows elastic scattering. Figure 4 presents the shape of the field $\varphi(x)$ in three different positions: before the interaction, after the interaction and on top of the potential. Figure 4 shows that the field moves with time without any noticeable changes in its shape, which is another indication of soliton-barrier elastic interaction.

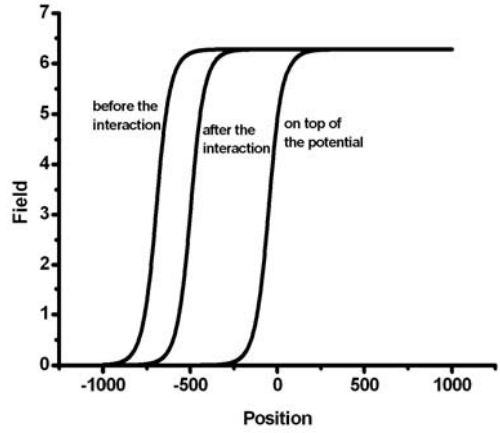

**Figure 4:** Situation of the field before, during and after the interaction

A soliton with initial velocity $u_i > u_c$ has enough energy for climbing the barrier, and passing over the potential. At the velocities $u_i \approx u_c$ soliton interacts with the potential slowly and spends more times near the barrier, but this situation is not a bound state.

The critical velocity can be found by sending a soliton with different initial speed and observing the final situation after interaction (falling back or getting over the potential). Simulations give the value $u_c = 0.3934479$ for the potential $V(x) = 0.5\exp(-4(x-1)^2)$ in the range of numerical precision. Figure 5 presents the trajectory of a soliton with different values for the initial velocity $u_i$ close to the critical velocity $u_c$.

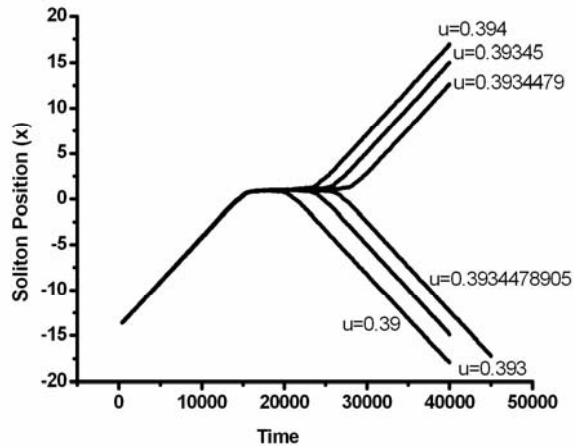

**Figure 5:** Trajectory of a soliton with different initial velocity $u_i \approx u_c$

Comparing the strength of the potential (here $V(x) = 0.5\exp(-4(x-1)^2)$) and the energy of the moving soliton shows that soliton can get over the barrier if its energy is more than the energy of a static soliton on top of the barrier. Thus the critical velocity is equal to the initial velocity of a soliton which has maximum energy of the barrier. For the above potential, maximum energy of a static soliton is: 9.5925 at x=1. Let us call this energy as critical energy ($E_c$). Solitons with energy $E_{max} < E_c$ can't pass the barrier. For example consider a soliton with initial velocity $u_i$=0.3934478. It has a peak energy of $E_{max}$=8.9966<$E_C$, therefore it reflects from the barrier. In other hand, a soliton with initial velocity $u_i = 0.3934479$ has the maximum energy: $E_{max}$=9.6022. Figure 4 shows that it can pass over the potential.

Figure 6 presents the final velocity of a moving soliton after interacting with the potential as a function of its initial velocity. For velocities smaller than $u_c$, soliton reflects back with a final negative velocity. Solitons with velocities above the $u_c$, transmit over the barrier. Small fluctuations appeared in large velocities are not bigger than the maximum of numerical error.

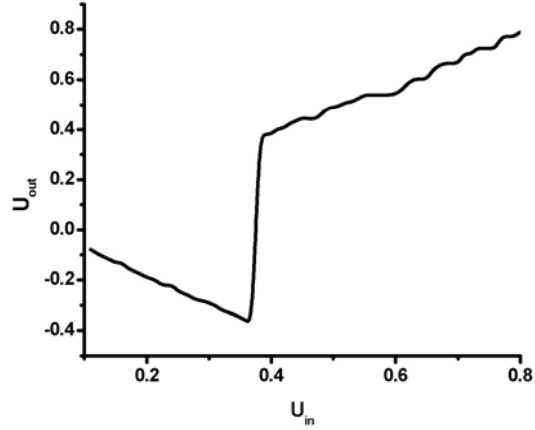

**Figure 6:** Final velocity of moving soliton after interacting with the potential $V(x) = 0.5\exp(-4(x-1)^2)$ respect to its initial velocity.

Several simulations for soliton-barrier (and also anti soliton-barrier) interaction were setup with different parameters for the potential and the soliton. All of them show elastic behaviour of soliton-barrier system.

**4. Interaction of a soliton with a potential well**

Suppose a particle moves toward a frictionless potential well. It falls in the well with increasing velocity and reaches the bottom of the well with its maximum speed. After that, it will climb the well with decreasing velocity and finally pass through the well. Its final velocity after the interaction is equal to its initial speed. We expect the same behaviour for the interaction of soliton-well system. Let us investigate this situation.

It is clear that $V(x) = a\exp(-b(x-c)^2)$ with $a < 0$ corresponds to a potential well. Like the case of the barrier, we can find the shape of the potential as seen by the soliton by plotting the total energy of a static soliton placed in different position (x). Figure 7 shows the results for the potential $V(x) = -0.5\exp(-4(x-1)^2)$ (solid line). The dashed line in figure 7 presents $E_0 + V(x)$, where $E_0$ is the energy of static soliton in the absence of the potential and V(x) is the potential function.

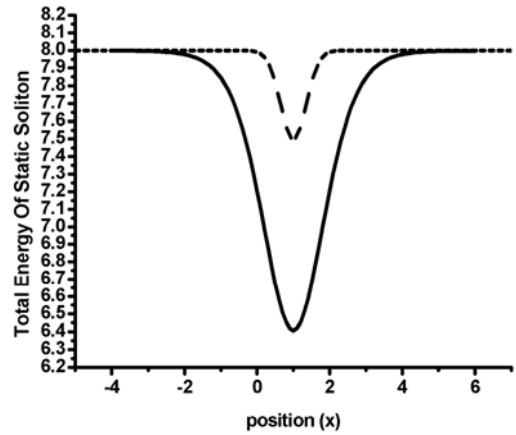

**Figure 7:** Total energy of a static soliton (solid line) placed in different position, under the effects of the potential $V(x) = -0.5\exp(-4(x-1)^2)$ and $E_0 + V(x)$ (dashed line).

Soliton-well interaction is a complicated interaction. Imagine a soliton which is located far from the potential, moves toward the well. It is expected that the soliton passes through the potential with any value for its initial velocity. But surprisingly it is not true!

In this case again, there exists a critical velocity ($u_c$) which separates two different kinds of soliton behaviour. A Soliton with initial speed above the $u_c$, transmits through the well and solitons with initial velocity lower than the $u_c$ fall into the well and is trapped by the potential. For the potential $V(x) = -0.5\exp(-4(x-1)^2)$, critical velocity is $u_c \approx 0.22470$ which has been found by numerical trial and error (see figure 8).

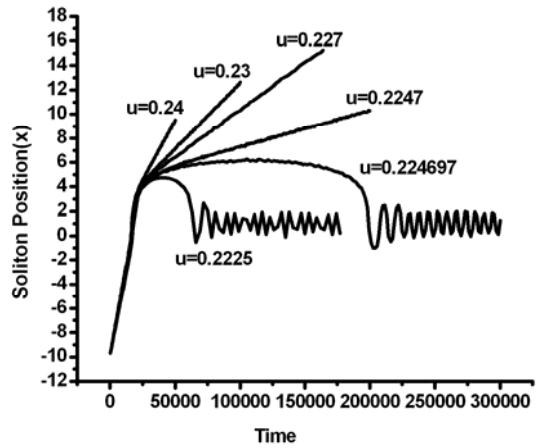

**Figure 8:** Trajectories of a soliton with different initial velocity during interaction with potential $V(x) = -0.5\exp(-4(x-1)^2)$

Consider a soliton with a large initial velocity. The energy of such a soliton is higher than the potential, so the soliton transmits through the well without any considerable effects. For this interaction, the final energy and velocity of the soliton is smaller than these quantities before the interaction. Also we can see a small shift in its trajectory (see figure9). Soliton has its maximum velocity near the well, so it stays a short period of time around the potential. This is the reason for the shift (which has been appeared in figure9) in soliton trajectory.

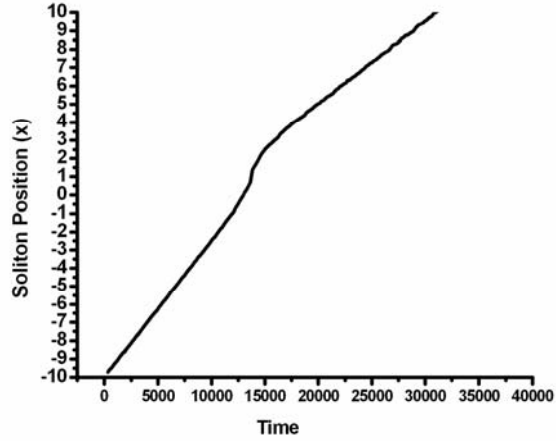

**Figure 9:** Small shift in the trajectory of a soliton around the location of the potential

A soliton with an initial speed close to the critical velocity shows an interesting behaviour. Consider a soliton with an initial velocity, a little bit above the $u_c$. It will stay near the well for a long time with a very slow motion, but gradually goes far from the well with an increasing speed and emits some energy. Simulations show that its final energy (velocity) is less than the initial energy (velocity). Clearly this situation can not be a bound state.

Now imagine a soliton with an initial speed lower than $u_c$, but very close to this value. It falls into the well and stays there oscillating. Soliton emits energy during oscillation. The amount of emitted energy per cycle decrease in time and the soliton-potential system approaches a stable situation (See the soliton trajectory with initial velocity $u_i = 0.2226$ or $u_i = 0.224697$ in figure 8). Figure 10 shows the evolution of the field $\varphi(x)$. Unlike the Soliton-barrier case, there is some deformation in the shape of the field because of soliton radiation during its evolution.

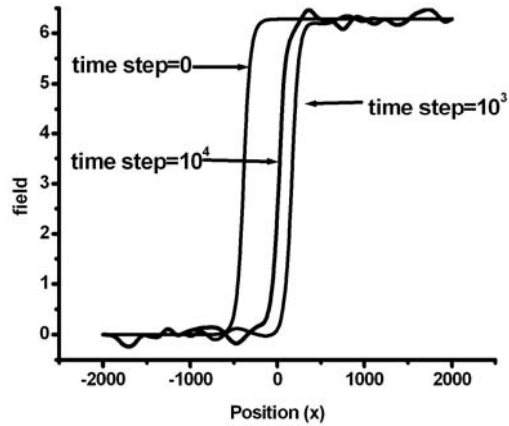

**Figure 10:** Time evolution of the field $\varphi(x)$ shows some defects come from energy radiation.

The most interesting behaviour is seen in some very narrow windows of initial velocity. Amazingly at some velocities smaller than $u_c$ the soliton reflects back or transmits over the potential while one would expect that soliton should trap in the potential. These narrow windows were found by scanning the initial velocity with small steps. Figure 11 shows the final velocity of a moving soliton with initial velocity of $u_{in}$ after interacting with the potential $V(x) = -0.5\exp(-4(x-1)^2)$. In figure 11, $U_{out} = 0$ shows trapped solitons, while $U_{out} > 0$ indicate soliton transmission and $U_{out} < 0$ show reflection. This is the same effect re-ported in [9] for non-topological solitons. Repulsion of the sine-Gordon solitons by a delta-well defect has been reported in [11] too. But it is noticeable that, figure 11 shows, solitons may transmit through the well at low velocities. For the potential $V(x) = -0.5\exp(-4(x-1)^2)$ reflection occurs in some velocities (For example $u_{in}$ =0.2130, 0.2195, 0.2240, 0.2250, 0.2340 and ….) and repulsion happen at $u_{in}$ = 0.2400, 0.2550, 0.2620, 0.2640.

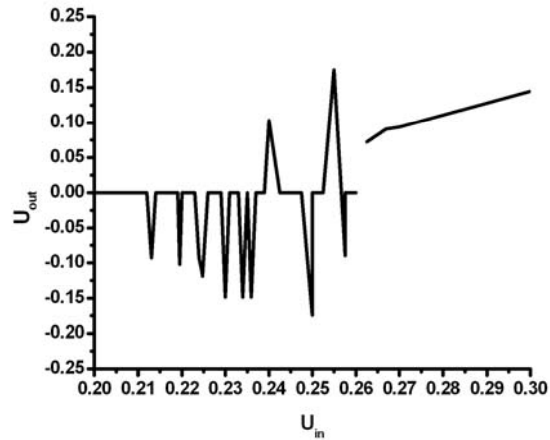

**Figure 11:** Final velocity of moving soliton after interacting with the potential $V(x) = -0.5\exp(-4(x-1)^2)$ respect to its initial velocity.

Several soliton-barrier simulations were setup with different parameters for the soliton and the potential. All the simulations demonstrated energy radiation during the interaction. Simulations show that the amount of radiated energy is a complicate function of soliton-well parameters, like soliton initial velocity and the potential parameters. These results are the same as what have been observed for the (2+1) dimensional baby Skyrme solitons in [10]. It is noticeable that the types of topological solitons under investigation in this paper and [10] are different. Moreover methods for introducing the potentials are very different in this paper and [10] too.

## 5. Conclusion

A soliton interacts with a potential barrier elastically, simply like a point particle. At low velocities it reflects back and with a high velocity climbs the barrier and transmits over the potential. Energy exchanges between soliton and the barrier during the interaction. The final speed of the soliton after the interaction is equal to its initial speed with a very good approximation. There exists a critical velocity $u_c$ which separates these two kinds of trajectories.

But for the case of soliton-well interaction, we observe interesting effects. A high speed soliton passes through the potential well and a low speed soliton becomes trapped in the well and oscillates there. In both situations the soliton emits energy, so its energy decreases in time to a stable state. Also at narrow windows of initial velocities, lower than the $u_c$, solitons may reflect back or transmit through the well.

**Acknowledgements**
I would like to thank W. J. Zakrzewski for inspiration and comments and also M. Sarbishaei, M. R. Garousi and A.R. Mokhtari for useful comments.